\documentclass[12pt,a4paper]{article}
\usepackage{amsmath}
\usepackage{amssymb}
\usepackage{bm}
\usepackage{graphicx}
\usepackage{amsbsy}
\usepackage{rotating}
\usepackage{latexsym}
\usepackage{pdflscape}

\newcommand{\pkg}[1]{{\normalfont\fontseries{b}\selectfont #1}}
\let\proglang=\textsf

\title{Long-range properties and data validity for hydrogeological time series:\\the case of the Paglia river}

\author{Marcel Ausloos\thanks{Corresponding author}
 \\ \textit{\small University of Leicester, School of Business}  \\ \textit{\small Leicester, LE11 7RH, United Kingdom }\\{\small Email:} \texttt{\small ma683@le.ac.uk}
  \\ \textit{\small Group
of Researchers for Applications of Physics in Economy and Sociology (GRAPES), }  \\ \textit{\small
  rue de la Belle Jardiniere 483, B-4031 Angleur, Belgium }\\{\small Email:} \texttt{\small marcel.ausloos@ulg.ac.be}
\and
Roy Cerqueti \\ \textit{\small University of Macerata, Department of Economics and Law, Macerata, I-62100.}\\{\small Email:} \texttt{\small roy.cerqueti@unimc.it}
\and
Claudio Lupi \\ \textit{\small University of Molise,  Department of Economics, Campobasso, I-86100.}\\{\small Email:} \texttt{\small lupi@unimol.it}
}
 
\begin{document}
\maketitle

\begin{abstract}
This paper explores a large collection of about 377,000
observations, spanning more than 20 years with a frequency of 30
minutes, of the streamflow of the Paglia river, in central Italy.
We analyze the long-term persistence properties of the series by
computing the Hurst exponent, not only in its original form but also
under an evolutionary point of view by analyzing the Hurst exponents
over a rolling windows basis. The methodological tool adopted for
the persistence is the detrended fluctuation analysis (DFA), which
is classically known as suitable for our purpose. As an ancillary
exploration, we implement a control on the data validity by
assessing if the data exhibit the regularity stated by Benford's
law. Results are interesting under different viewpoints. 
First, we show that the Paglia river streamflow
exhibits periodicities which broadly suggest 
the existence of some common behavior with El Ni\~no and the North Atlantic
Oscillations: this specifically points to a (not necessarily direct) 
effect of these oceanic
phenomena on the hydrogeological equilibria of very far geographical
zones: however, such an hypothesis needs further analyses to be validated.
Second, the series of streamflows shows an
antipersistent behavior. Third, data are not consistent with
Benford's law: this suggests that the measurement criteria should be
opportunely revised. Fourth, the streamflow distribution is well
approximated by a discrete generalized Beta distribution: this 
is well in accordance with the measured streamflows being the outcome
of a complex system. 

\vspace{1cm} \noindent \textbf{Keywords:} River streamflow, Hurst
exponent, Benford's law, Detrended fluctuation analysis, Discrete
generalized Beta distribution.
\end{abstract}

\section{Introduction}

Hydrogeological paths exhibit often a complex behavior. Such
irregularities are driven by the presence of several factors of
uncertainty in the climate dynamics. River streamflows represent 
paradigmatic examples of this evidence. In fact, human activities and
natural events drive the fluctuations, sometimes with catastrophic effects,
of such phenomena.

This paper elaborates some crucial observations from these premises.
In particular, we develop an analysis of the streamflow of the
Paglia river, a major tributary of Tiber, whose watercourse is
entirely localized in Central Italy. Data are the official
measurements of the streamflow of the river at Ponte dell'Adunata, near
Orvieto, a historical town in Umbria region. The time interval
covered by the measurements is January 1st, 1992 (12:00am) -- May 13,
2014 (h11:30pm), and the periodicity of the observations is 30
minutes. However, the pervasive presence of missing values
in the first part of the series prevented us from using the 
data prior to November 2, 1992 (h12:30pm).

We specifically focus on the so-called \emph{long run dependence
property} of the series, which gives a thoughtful view of the
behavior of the autocorrelation function. Such a statistical
property can be suitably employed for making forecasts on the
future evolution of the streamflow. The perspective presented here
is in line with the original, path-breaking study by [\cite{hurst1951long}],
who explored the long-run dependence of the runoffs of the Nile
river. The analysis is based on the assessment of the value of the
constant $H \in (0,1)$ --- the so-called \emph{Hurst exponent} ---
which represents the rate of decay of the autocorrelation function
of the series as a function of the time-lag. If $H>1/2$, the
series is said to present a persistent behavior, which basically
means that the history of the past will ``statistically'' repeat in
the future; $H<1/2$ stands for antipersistence, the opposite of
persistence; $H=1/2$ is the pure random case.

Since its inception, several contributions dealing with the
assessment of the Hurst exponent in a number of very
different contexts have appeared in the literature;  a complete list is too long to
be mentioned here.

In detecting long-run dependence, river discharges are of particular
interest. In fact, they offer very specific time series with
peculiar features. First of all, they have periodicities,
with different periods. Such a property is due to the strong
relationship between the river streamflow dynamics, the weather and the
precipitation, but also to the effects of the global warming over
the hydrogeological systems of the planet Earth 
[\cite{hurst1951long}-
\cite{hajian2010multifractal}].

Under the point of view of the human activities, it is self-evident
how the fluctuations of the streamflow of any river affect (positively
or negatively) the socio-economic system of the surrounding
areas. In this respect, the Paglia river and the municipality of Orvieto represent very
interesting cases to treat. In fact, there is a high level of
hydrogeological risk related to Paglia's floods in the Orvieto area:
the last important flood took place on November 12, 2012, causing
injuries and huge damages to the local economy.
An effective plan to deal with this recurring phenomenon should
necessarily move from a deep knowledge of the dynamics of the river
and a forecast of when and how a flood will take
place in the future.

From a purely technical viewpoint, we proceed by applying
the \emph{Detrended Fluctuation Analysis} (DFA) ---
introduced by [\cite{peng1994mosaic}] --- which is classically acknowledged to be a
powerful tool for the estimation of the Hurst exponent,
whence long term persistence effects, mainly in the
presence of non-stationarities. The appeal of the DFA lies in its
conceptualization. Indeed, the underlying theoretical framework can
be found in the theory of random walks [\cite{shlesinger1987levy,
ben2000diffusion}]. In such a context, time series
are opportunely aggregated. This reduces the noise level due to
biases in measurements.

In the context of DFA applications, it is worth
mentioning [\cite{ivanovaausloos1999}-
\cite{huetal2001}].

As an ancillary analysis, we also consider the ``validity'' of the
data measured at Ponte dell'Adunata. We do not infirm its interest,
of course, but we wish to point some possible defects impairing a
finer analysis as that presented here. In fact, the subsequent
discussion and conclusions might be the first of this sort in the
present context. For this purpose, and in accordance with
[\cite{nigrinimiller2007}] observations of river streamflows in the USA, we
check if data fit Benford's law [\cite{benford1938}].

In fact, such an empirical rule (on the logarithmic
distribution of the first digit in lengthy data) has been found to
hold in a wide spectrum of cases; see, e.g.,  (and the
references therein] [\cite{judgeschechter2009}-
\cite{ausloos2016regularities}]. When violated, the
presence of some sort of data manipulation or mistakes in data
collection has to be debated.

The rest of the paper is organized as follows: in the next Section (Sect. \ref{sec:datamethod})
we introduce  the dataset and the methodologies. In Section
\ref{sec:results} we present  and discuss the results. In the last Section (Sect.
\ref{sec:conclusion}), we offer  some concluding remarks.

\section{Data and methodologies}\label{sec:datamethod}
\begin{figure}
\begin{center}
\includegraphics[scale=.65]{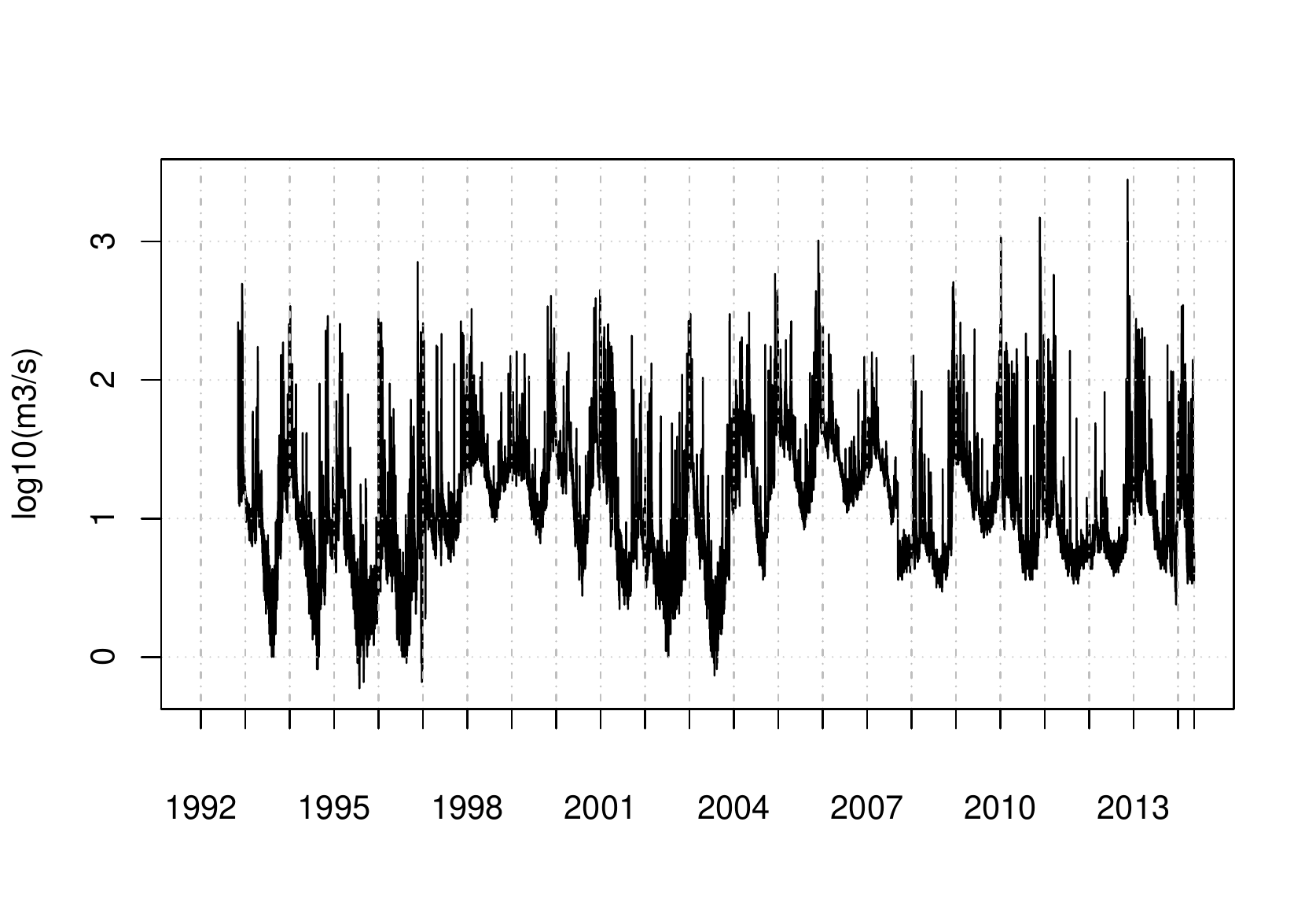}
\end{center}
\vspace{-1cm} \caption{\label{fig:paglia_series_log} Streamflow (expressed in $\log_{10}\left(m^3/s\right)$) of the Paglia river measured at every half hour between November 2, 1992 and May 13, 2014. Ticks on the x-axis identify the beginning of the years, except the the last tick that indicates the last available observation.}
\end{figure}

The Paglia river is an important right-side tributary of Tiber, 
the third longest river in Italy.
It is about 86km-long, with many tributaries along its course.
It is characterized by a very variable streamflow 
(see Table~\ref{tab:descr.stats} and Figure~\ref{fig:paglia_series_log})
which gives Paglia a crucial role in determining the floods of Tiber.
Data, obtained from an hydrometer station run by the Umbria regional authority,
measure the streamflow (expressed in $m^3/s$) of the Paglia river at Ponte
dell'Adunata, just downstream of the confluence of Chiani river, a 42km-long
left-side tributary of Paglia.

Because of the presence of many long measurement interruptions
during the first ten months of 1992, in our analysis we consider
measurements recorded at each half hour starting on
November 2, 1992 (h12:30pm), until May 13, 2014 (h11:30pm). There 
are still a few missing values in the selected period 
(in 7 occurrences the interruption of the
measurements lasted for more than 2 days). Such missing values have
been linearly interpolated on the series expressed in
logarithms.\footnote{All computations have been carried out using
\proglang{R} ver. 3.3.1 [\cite{R}] and packages
\pkg{benford.analysis}[ \cite{benford.analysis}], \pkg{fractal}
[\cite{fractal}], \pkg{ggplot2} [\cite{ggplot2}], \pkg{imputeTS}
[\cite{imputeTS}], \pkg{moments} [\cite{moments}], \pkg{psd} [\cite{barbourparker2014}],
\pkg{xts} [\cite{xts}].} The resulting series (plotted in logs in
Figure~\ref{fig:paglia_series_log}) includes 377,399 observations.

\begin{table}
\begin{center}
\begin{tabular}{cccccc}
\hline \hline
Min.     & 1st Q.   & Median   & Mean  & 3rd Q. & Max. \\
0.59     & 6.29     & 12.31    & 20.53 & 23.93  & 2794.00 \\ \hline
$\sigma$ & Skewness & Kurtosis & Date of Max. & Date of Min. & Data points\\
34.63    & 20.81    & 989.48   & 2012/11/12   & 1995/07/28   & 377399 \\ \hline \hline
\end{tabular}
\end{center}
\vspace{-0.5cm}\caption{Main descriptive statistics of the Paglia river streamflow $(m^3/s)$ as measured at Ponte dell'Adunata, near Orvieto (November 2, 1992 -- May 13, 2014).}
\label{tab:descr.stats}
\end{table}

Data have been analyzed under several perspectives: 
first, we carry out a graphical analysis of the main features of the streamflows.

Then, we estimate the spectrum of the daily averages, in order to assess
the presence of seasonal and non-seasonal periodicities.

For what concerns the estimation of the Hurst exponent, we perform a detrended
fluctuation analysis (DFA) over 90-day long rolling windows. 
At each iteration the window is moved forward by a 1-day step. 
For a more intuitive visualization, when needed, windows will be 
numbered according to a chronological criterion. Specifically, 
we start from window 1 --- the one at the beginning of the sample period. Then, we
shift the window by one day and add 1 to the windows counter at any shift. According to
this mechanism, window 1 is the one spanning days 1--90,
window 2 spans days 2-91, window 3 insists on days 3-92, and so on. 
The Hurst exponent, $H$, is estimated in each window,
resulting in a series of 7,773 estimated values of the Hurst
exponent. In order to check the sensitivity of the results on the window
length, we repeat the same analysis using 180 and 400-day long windows.

We also carry out a check for ``data validity''. For this purpose,
we adopt the point of view of [\cite{nigrinimiller2007}], who
suggest that streamflow statistics of US rivers are broadly
consistent with Benford's law. For this reason we check if the data
gathered for the Paglia river are also consistent with Benford's law.
This part of the analysis is carried out on the observed data only,
without any imputation of missing values. In particular we carry out
a first-two digits test, which is more informative than the
combination of both the first digit and the second digit tests;
 see e.g., especially Chapter 4 in [\cite{nigrini2012}].

Finally, we check if the observed data approximately follow 
a discrete generalized Beta distribution (DGBD), typical
of the output of complex systems [\cite{naumiscocho2008}].


\section{Results and discussion}\label{sec:results}

\begin{figure}
\begin{center}
\includegraphics[scale=.65]{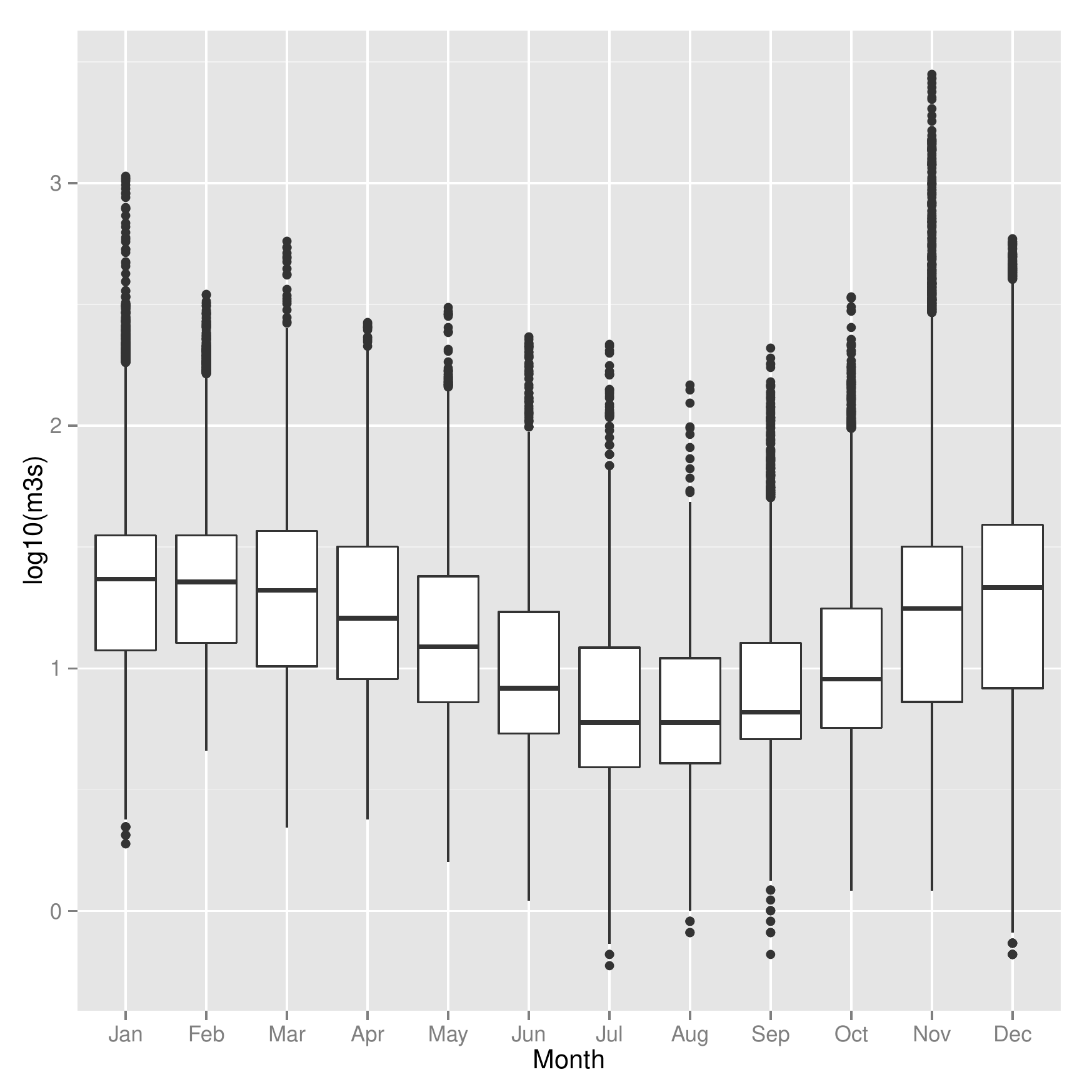}
\end{center}
\vspace{-1cm} \caption{\label{fig:flow_month} Distribution of the
streamflow ($\log_{10} \left(m^3/s\right)$) of the Paglia river by month.}
\end{figure}

The streamflow series expressed in logarithms shows several features of the
Paglia river fluctuations (see Figure~\ref{fig:paglia_series_log}).
However, more information can be gathered by looking at the basic
monthly statistics of the data: it is evident that the
series exhibits some periodicity, visually two regimes: a stationary one,
representing the ``normal'' streamflow of the Paglia river, and a peaked one,
which captures the cases of floods. 
This is nicely exemplified by displaying the monthly
distribution of the water flow as provided in
Figure~\ref{fig:flow_month}: significant variations are apparent.
Indeed, the streamflow of the river is more powerful in winter
than in summer, and this is in line with the standard precipitation
cycle in Italy, which exhibits a periodicity associated to the
seasons. However, the highest streamflow peaks are all concentrated
in November, even if the median value of the streamflow in that month is smaller than
in the other winter months. This observation suggests that 
the maximum intensity of precipitations
in the area is historically located in November, despite this month
not being the most rainy one.

\begin{figure}
\begin{center}
\includegraphics[scale=.65]{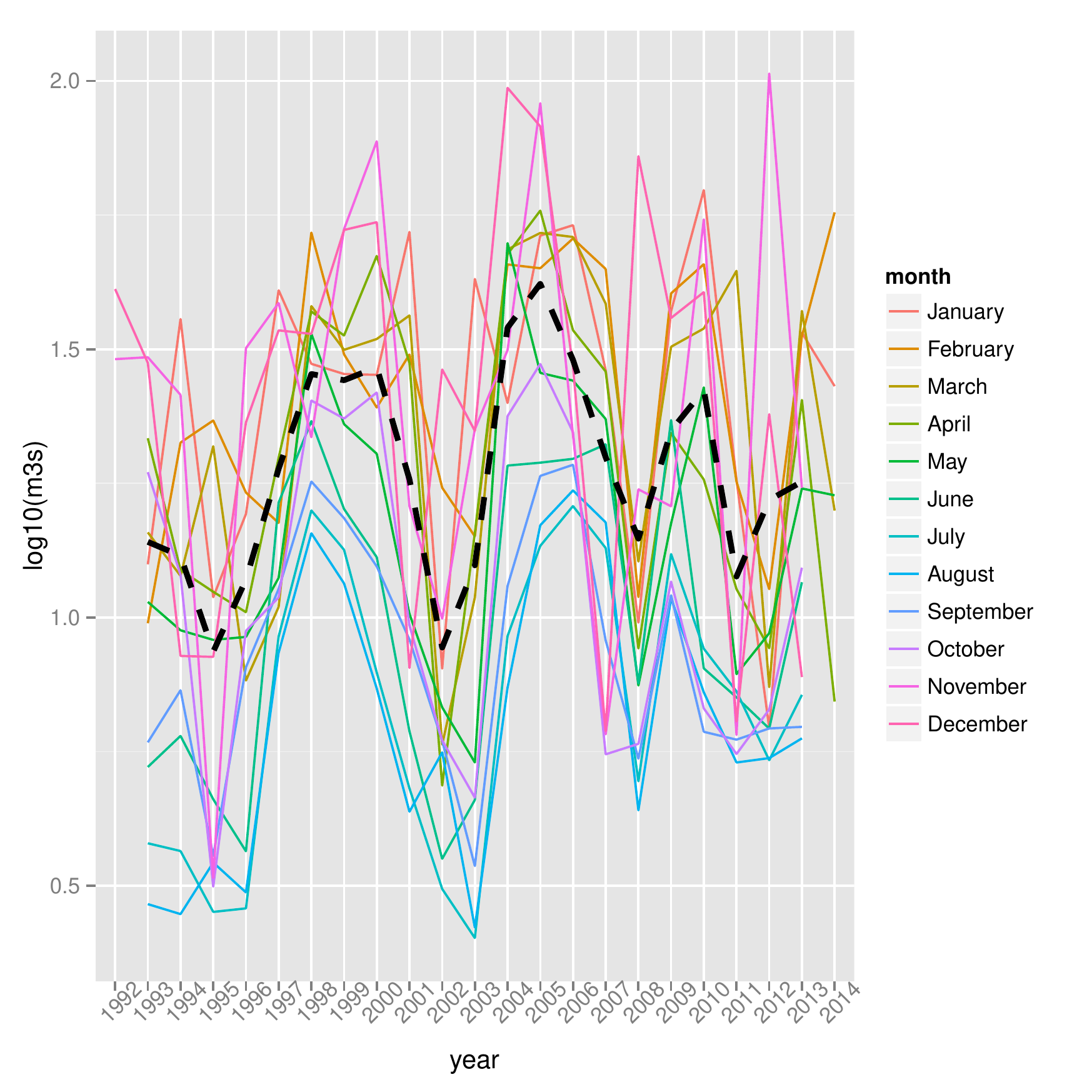}
\end{center}
\vspace{-1cm} \caption{\label{fig:monthly_avg} Evolution of monthly
average streamflow ($m^3/s$) of the Paglia river. The black dashed line
represents the yearly average.}
\end{figure}

\begin{figure}
\begin{center}
\includegraphics[scale=.65]{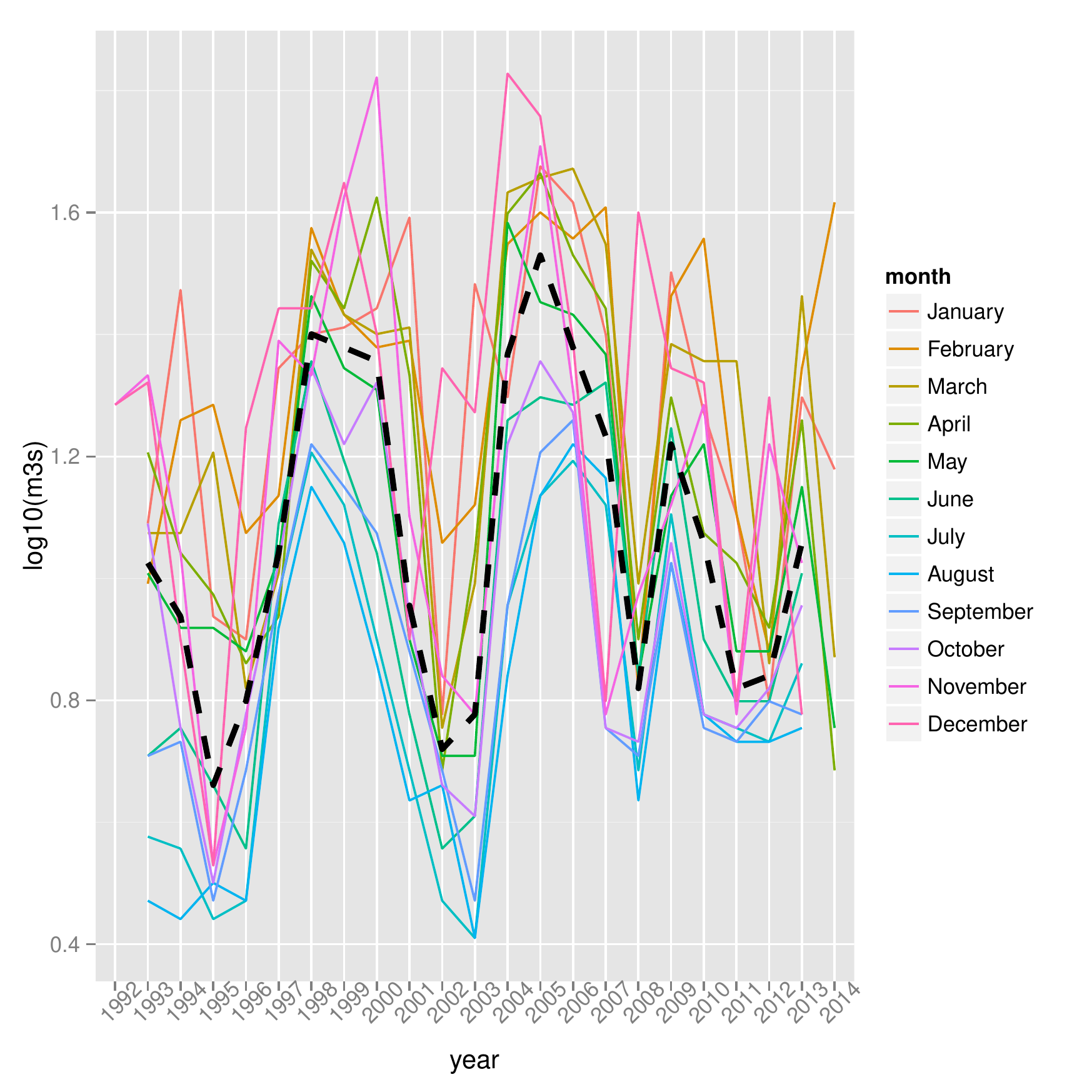}
\end{center}
\vspace{-1cm} \caption{\label{fig:monthly_med} Evolution of monthly
median streamflow ($m^3/s$) of the Paglia river. The black dashed line
represents the yearly median.}
\end{figure}

The evolution of the monthly averages is plotted in
Figure~\ref{fig:monthly_avg}. If the median streamflow is  considered
instead of the average, the overall picture remains substantially
unchanged and is possibly even clearer (Figure~\ref{fig:monthly_med}). 
A cyclical pattern is apparent
from the data, with a cycle whose period is about 5--7 years.
This result suggests the existence of a relationship between 
rivers streamflow fluctuations and the periodicity of the oscillations of
El Ni\~{n}o, which is recurrent on irregular intervals with an
average period of 3--4 years;  
see, e.g., [\cite{cane1986experimental,eltahir1996nino}]. 
However, it is worth to point out that the observed periodicities 
of these different phenomena do not perfectly match, 
indicating that further investigations are needed for a confirmation
of this hypothesis. Furthermore, such a
periodicity is also broadly in accordance with that of the North Atlantic
Oscillations, whose fluctuations show a fairly regular periodic behavior
over a time interval of 6--8 years [\cite{colletteausloos2004}].

\begin{figure}
\begin{center}
\includegraphics[scale=.65]{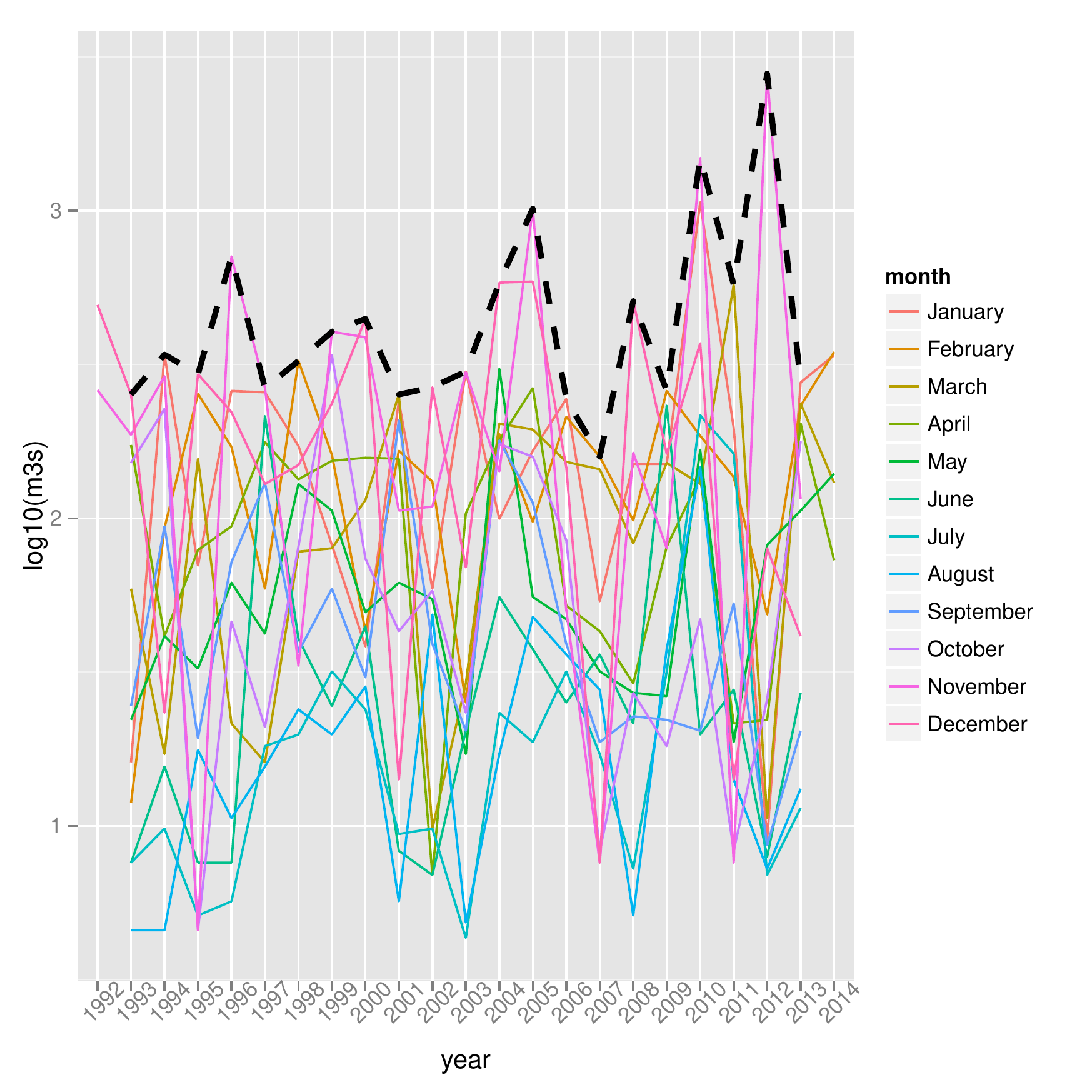}
\end{center}
\vspace{-1cm} \caption{\label{fig:monthly_max} Evolution of monthly
maxima ($m^3/s$) of the Paglia river. The black dashed line represents
the yearly maximum.}
\end{figure}

Equally interesting is the
evolution of the monthly maxima, reported in
Figure~\ref{fig:monthly_max}, where a positive trend seems to exist  in the recent annual maxima of
the streamflow. This means that the intensity of the streamflow
maxima has grown with respect to time in the last few years. This
finding is in line with the overall climate change, which represents
a severe and debated concern.

\begin{figure}
\begin{center}
\includegraphics[scale=.65]{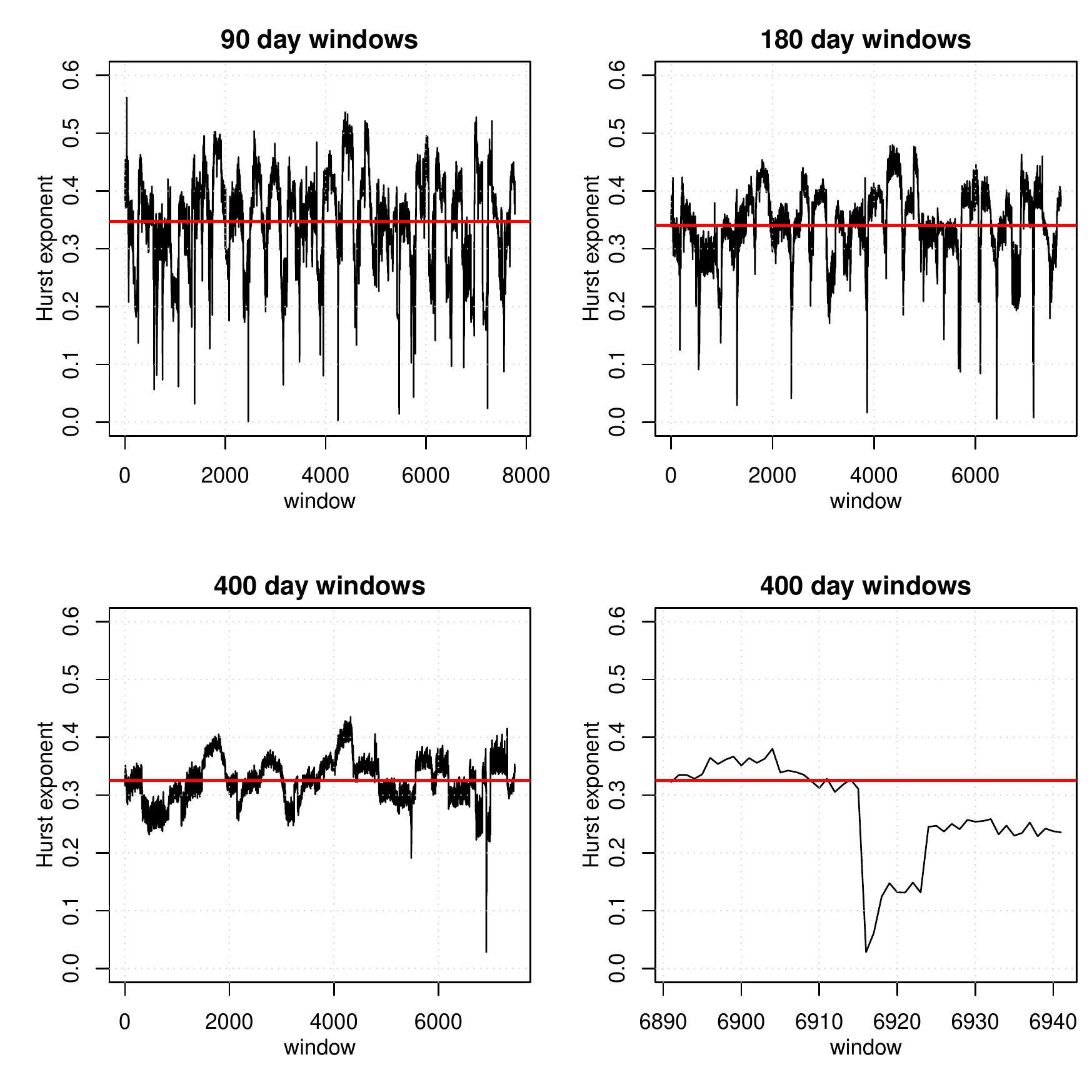}
\end{center}
\vspace{-1cm} \caption{\label{fig:hurst1_series} Estimated Hurst exponent over 90, 180, and 400-day long rolling windows. At each iteration the window is moved forward by a 1~day step. On the x-axis is reported chronological order given to windows. The horizontal red line is the average estimate of the Hurst exponent over all the windows. The fourth (lower-right) panel focuses on a subset of windows centred on the negative peak of the 400-day long windows estimates, corresponding to the first window containing November 12, 2012.}
\end{figure}

The rolling estimates of the Hurst exponent are plotted in 
Figure~\ref{fig:hurst1_series}: 
at first sight this figure seems to suggest the presence of frequent large
oscillations of the estimated values of the Hurst exponent in
adjacent windows. In fact, this is not the case and this impression is 
mainly due to the high density of data points along the x-axis in the graph. 
A closer examination of the estimated values of the Hurst exponent reveals 
that there are few relatively large oscillations, 
related to an extreme event entering (or leaving) 
the estimation window. This effect is clearly visible in
the fourth panel of Figure~\ref{fig:hurst1_series} that
focuses on a subset of windows in the neighborhood of the negative
spike depicted in the 400-day long windows panel.
Here it can be seen that the value of the Hurst exponent is fairly stable until the
data of the flood of November 12, 2012 enter the estimation window: in 
coincidence with the new data entering the window, the
estimated value of the Hurst exponent drops.

When considering 90-day long windows, 90\% of the variations of the estimates
across adjacent windows belong to the interval $(-0.043, 0.046)$ and about 3.6\% of them
are larger than 0.05 in absolute value, whereas only 0.4\% are
larger than 0.1 (less than one third of the average value). 
Furthermore, 90\% of the differences across 
adjacent 180-day long windows belong to the interval $(-0.032, 0.034)$ and less than 0.1\%
of the variations are larger than 0.1 in absolute value: when
400-day windows are considered, these values become $(-0.023, 0.022)$ and 0.03\%, respectively 
(see Table~\ref{tab:descr_H}). Note also that the estimated values are more 
stable for longer windows, as expected.

It is interesting also to observe that the average Hurst exponent over the rolling windows is about
$1/3$ (see again Table~\ref{tab:descr_H} and Figure~\ref{fig:hurst1_series}). 
This value means that the series of the Paglia river streamflows is highly antipersistent.
By analogy with financial econometrics based on DFA and time evolution of the Hurst exponent, this
observation should lead to constructive risk assessment measures, but an application or discussion
of this forecasting process falls outside the framework of the present paper. 
Uncertainty associated with the estimates is small, as can be observed from Table~\ref{tab:descr_H}.

\begin{sidewaystable}
\begin{center}
\begin{tabular}{lrrrrrrrrr}
\hline \hline
 & \multicolumn{3}{c}{90-day long windows} & \multicolumn{3}{c}{180-day long windows} & \multicolumn{3}{c}{400-day long windows}\\ \hline
\multicolumn{1}{r}{} & $H$ & $\sigma_H$ & $\Delta H$ & $H$ & $\sigma_H$ & $\Delta H$ & $H$ & $\sigma_H$ & $\Delta H$ \\ \hline
Min. & 0.001 & 0.006 & -0.230 & 0.006 & 0.008 & -0.204 & 0.028  & 0.019  & -0.283\\ 
1st Q. & 0.293 & 0.029 & -0.015 & 0.308 & 0.035 & -0.011 & 0.300  & 0.036 & -0.008\\
Median & 0.357 & 0.040 & 0.000 & 0.343 & 0.041 & 0.000 & 0.324  & 0.042 & 0.000\\ 
Mean & 0.347 & 0.040 & 0.000 & 0.340 & 0.042 & 0.000 & 0.325 & 0.042 & 0.000\\ 
3rd Q. & 0.406 & 0.049 & 0.015 & 0.383 & 0.048 & 0.011 & 0.353 & 0.047 & 0.008\\ 
Max & 0.562 & 0.201 & 0.195 & 0.480 & 0.204 & 0.148 & 0.436 &  & 0.114\\ 
$\sigma$ & 0.081 & 0.017 & 0.029 & 0.060 & 0.014 & 0.022 & 0.038 & 0.139 & 0.014\\
Skewness & -0.437 & 1.921 & -0.355 & -0.749 & 2.582 & -0.659 & -0.239 & 3.083 & -0.973\\
Kurtosis & 2.880 & 12.851 & 8.585 & 4.620 & 24.185 & 10.940 & 3.795 & 21.881 & 26.402\\ 
90\% low & 0.205 & 0.015 & -0.043 & 0.253 & 0.023 & -0.032 & 0.258 & 0.029 & -0.023\\ 
90\% up & 0.475 & 0.060 & 0.046 & 0.444 & 0.058 & 0.034 & 0.384 & 0.053 & 0.022\\ 
\%$|\Delta H|>0.05$  &   &   &  3.551  &   &   &  1.093 &  &  & 0.188\\  
\%$|\Delta H|>0.075$ &   &   &  0.849  &   &   &  0.208 &  &  & 0.054\\  
\%$|\Delta H|>0.1$   &   &   &  0.386  &   &   &  0.091 &  &  & 0.027\\  
\hline \hline
\end{tabular}
\end{center}
\vspace{-0.5cm}\caption{Descriptive statistics of the estimates of the Hurst exponent ($H$), of its standard error ($\sigma_H$), and of its variations in adjacent windows ($\Delta H$). ``90\% low'' and ``90\% up'' describe the smallest intervals including 90\% of the estimates. \%$|\Delta H|>x$ denotes the percentage of variations in adjacent windows larger than $x$ in absolute value.}
\label{tab:descr_H}
\end{sidewaystable}

\begin{figure}
\begin{center}
\includegraphics[scale=.65]{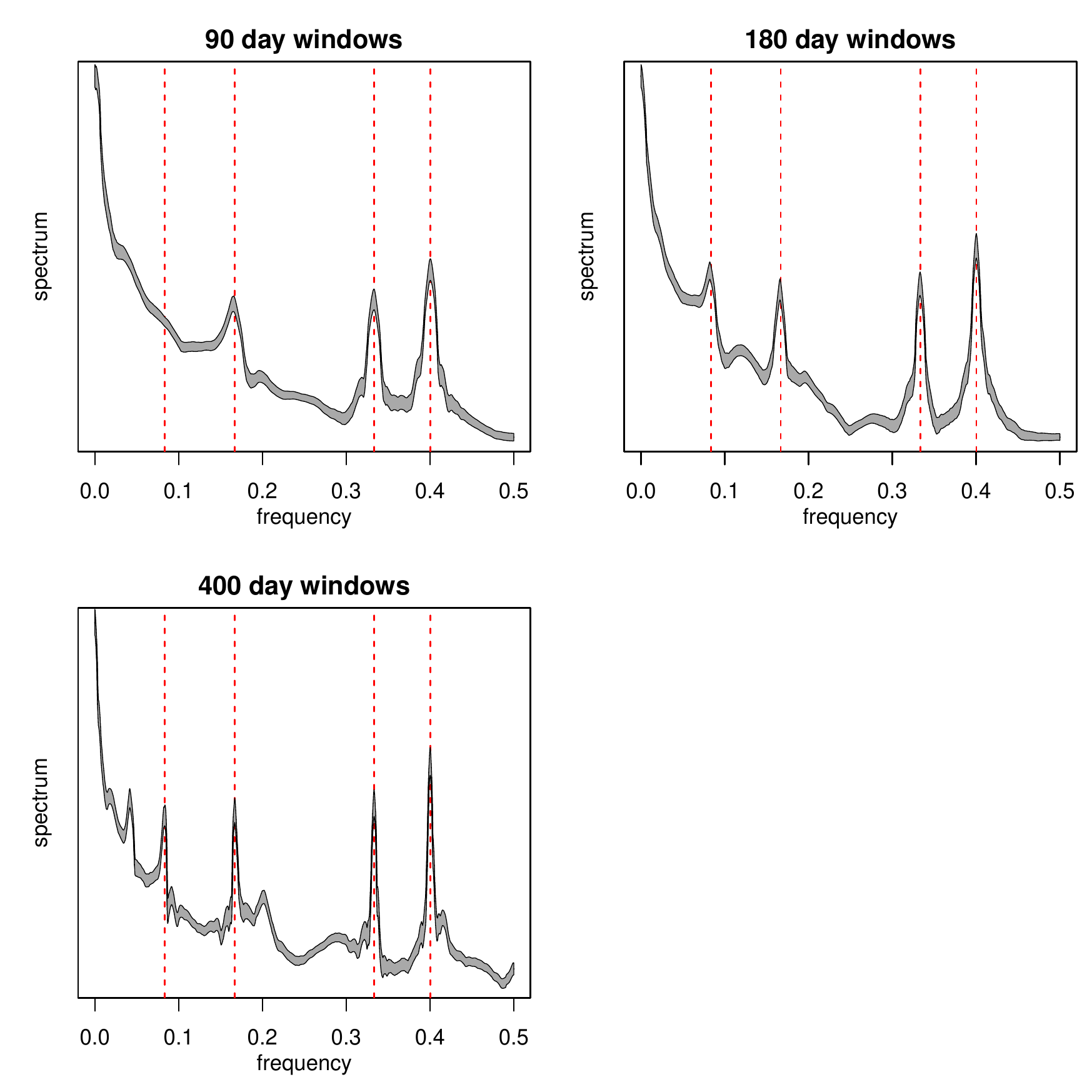}
\end{center}
\vspace{-1cm} \caption{\label{fig:hurst_spectra} Estimated spectra of the series of the estimated Hurst exponents. The grey areas represent spectral uncertainties for 95\% coverage probability. Peaks (from left to right) correspond to periodicities of about 12,  6,  3, and 2.5 days, respectively.}
\end{figure}

The estimates of the Hurst exponent seem to show a cyclical pattern, too.
For this reason we carry out a spectral analysis of the series of the
estimated Hurst exponents (see Figure~\ref{fig:hurst_spectra})
using an adaptive, sine multitaper power spectral density
estimation method [\cite{barbourparker2014}].

\begin{figure}
\begin{center}
\includegraphics[scale=.65]{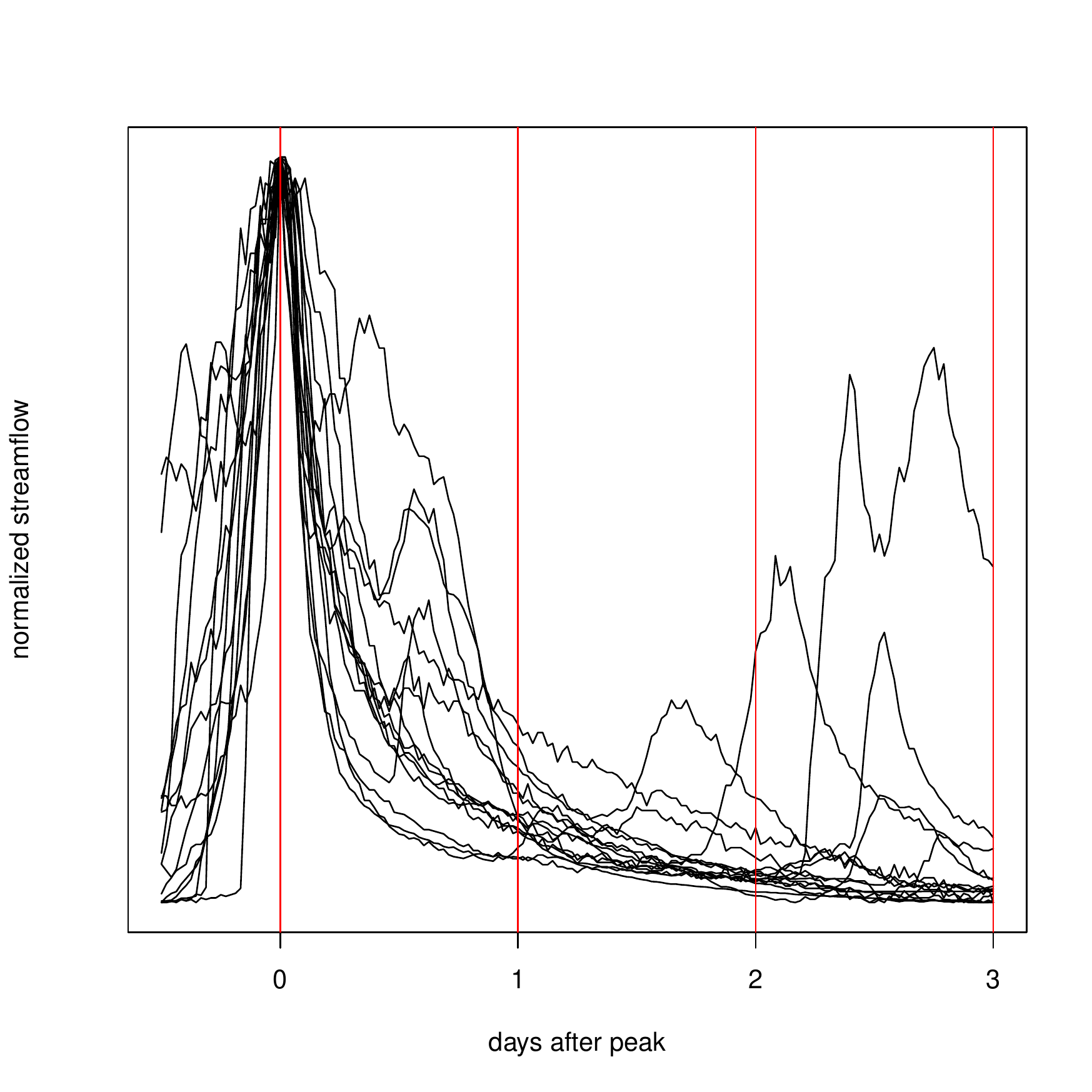}
\end{center}
\vspace{-1cm} \caption{\label{fig:days_after_peak_norm} Comparison of the 
streamflow dynamics in the neighborhood of a number of peaks. The x-axis represent
the distance from the peak expressed in days. The y-axis is the streamflow, normalized 
to have the same range across peaks.}
\end{figure}

We are not now in the position to offer a definitive answer to this
phenomenon. The apparent periodicity of the estimated Hurst exponent 
is probably influenced by the intrinsic features of the river streamflow:
in particular, we observe that 2--3 days is the time interval required
for the streamflow to return to the ``normal'' level preceding a peak, 
in the absence of further atmospheric phenomena. This is well illustrated 
in Figure~\ref{fig:days_after_peak_norm}, where the dynamics of
the streamflow in the neighborhood of a number of peaks is compared: 
for ease of comparison, the peak heights are normalized to have the same range. 
We conjecture that the nonlinear, fairly recurrent, dynamics 
highlighted in Figure~\ref{fig:days_after_peak_norm}
may lay behind some of the observed periodicities of the Hurst exponent.
This is not the only possible explanation, of course, and we
acknowledge that the influence of several factors affecting the estimation 
of the Hurst exponent though DFA may be large and such to determine 
possible periodicities in the rolling estimates. 
To the best of our knowledge, a list of such factors should necessarily include
the size of the time window, the number of data points in the
window, the position of the windows on the time interval, 
the presence of heteroskedasticity in the original data.  
The literature lists important contributions dealing with
estimation problems of the Hurst exponent and proposing methods
beyond the DFA [\cite{huetal2001,chenetal2002}] or
with different tools like moving average techniques 
[\cite{vandewalleausloos1998,vandewalleetal1999}]. 
Although this is beyond the scope of the present paper, 
the observed behavior of the spectrum of the Hurst exponent
suggests to carry out further investigations on the estimation
procedure of the Hurst exponent, possibly along the lines advocated 
in the contributions mentioned above.
At any rate, the order of magnitude of the Hurst exponent given by the DFA is incontestable.

\begin{figure}
\begin{center}
\includegraphics[scale=.65]{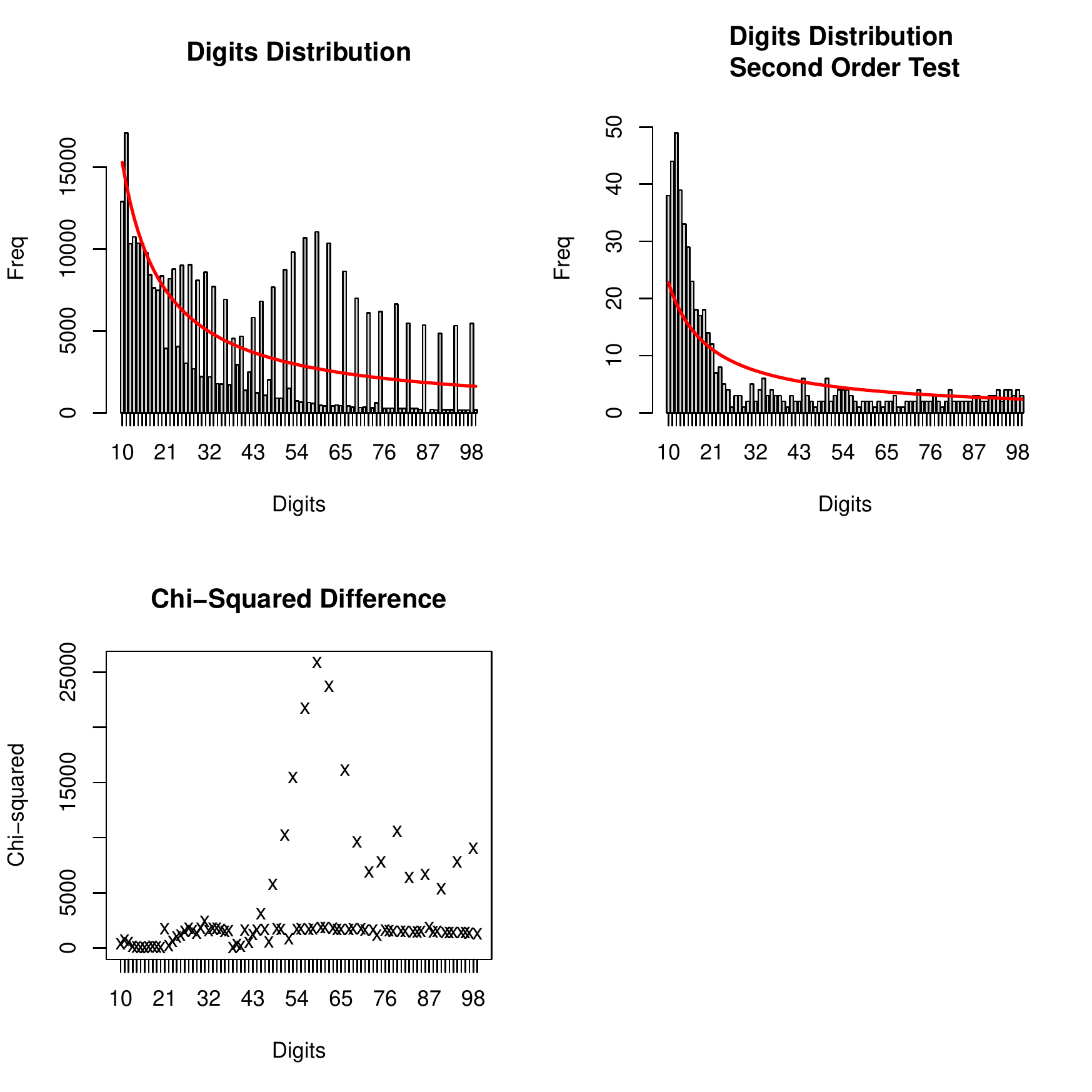}
\end{center}
\vspace{-1cm} \caption{\label{fig:benford} Benford's law first-two digits tests carried out
on streamflow data ($m^3/s$). Histograms represent the actual data distribution; the red solid curve is the theoretical Benford distribution.}
\end{figure}

We want now to focus on a potential weakness in the data recording process.
The validity check of the data, implemented through the
consistency with Benford's law, leads us to observe a substantial failure:
data seem to violate Benford's law (see Figure~\ref{fig:benford}).
Recall that Benford's  law (BL) describes the frequency
distribution of leading digits in  large data sets.
In particular, Benford's law on the first digit (BL1)
[\cite{newcomb1881,benford1938}] states that the distribution
of the first digit  is more concentrated on smaller values:
the digit "1" has the highest frequency, "9" the lowest frequency.
The first digit  distribution follows a logarithmic law:
\begin{equation}
P(d) = \log_{10} \left(1 + \frac{1}{d} \right),\qquad d=1,2,\dots ,9,  \label{BL}
\end{equation}
where $P(d)$ is the probability that the  first digit is equal
to $d$ in the data set; $\log_{10}$ being the logarithm in base 10.

\begin{figure}
\begin{center}
\includegraphics[scale=.65]{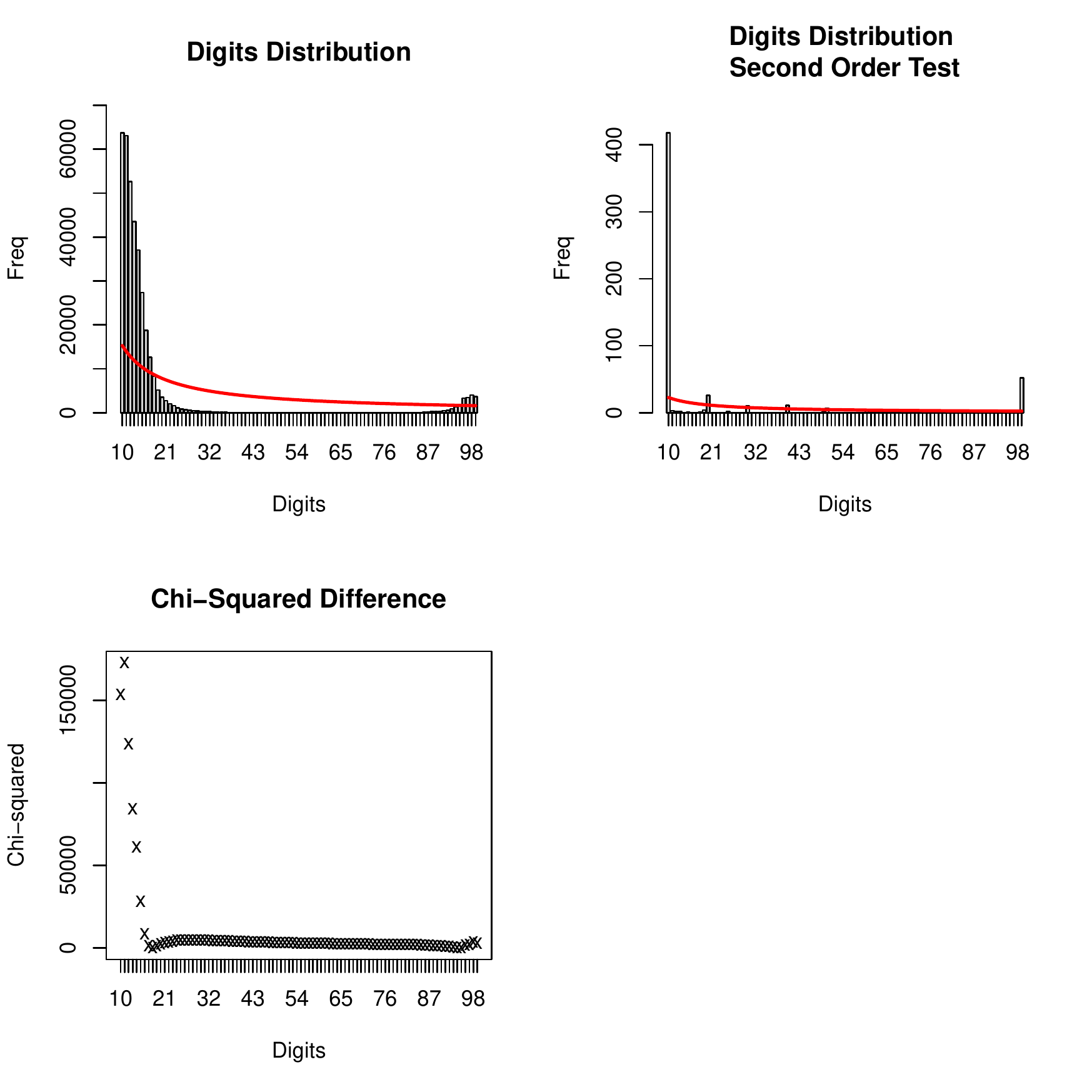}
\end{center}
\vspace{-1cm} \caption{\label{fig:benford_height} Benford's law first-two digits tests carried out on measured water heights ($cm$). Histograms represent the actual data distribution; the red solid curve is the theoretical Benford distribution.}
\end{figure}

There is a reasonable interpretation of the outcome of the
inconsistency of the data with the regularity imposed by
Benford's law. Flows ($m^3/s$) are computed from
water height, which is the really monitored quantity. However, water
height is recorded with one centimeter precision, so that many
repeated values are possible. In
fact, the two most common values are repeated more than 10,000 times
(see Table~\ref{tab:frequencies}). The analysis repeated on
water heights (measured in centimeters) reveals that the distribution
of water heights does not conform to Benford's law either (see
Figure~\ref{fig:benford_height}).
In fact, this is practically annoying when looking for streamflow
fluctuations in predictive analyses.

\begin{figure}
\begin{center}
\includegraphics[scale=.65]{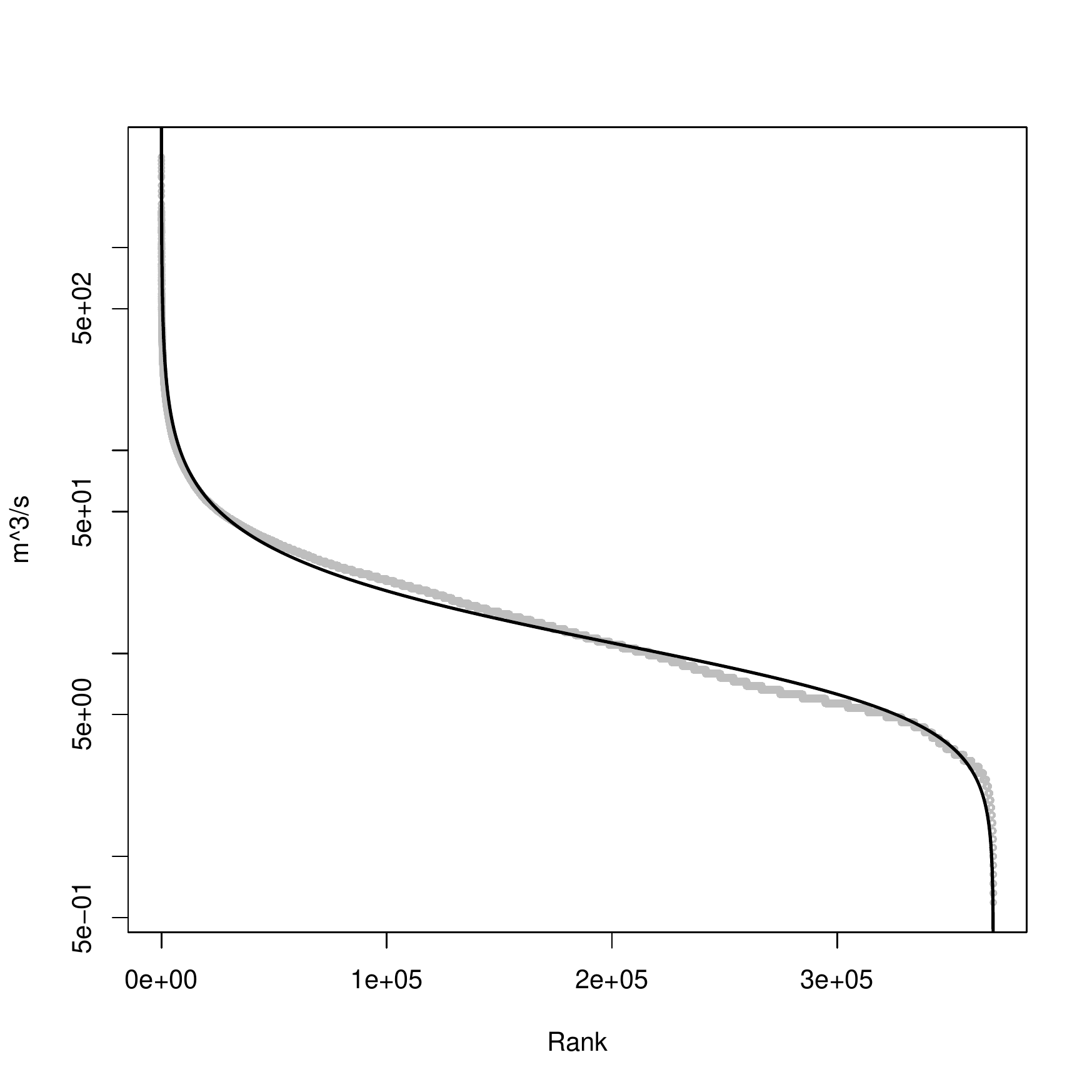}
\end{center}
\vspace{-1cm} \caption{\label{fig:log_rank} Rank-size relation of the Paglia river streamflow (data points are represented in grey). The continuous curve is the fit of a discrete generalized Beta distribution with $(A, a, b, R^2) = (163.117, 0.596, 0.382, 0.986$).}
\end{figure}

Finally, looking for other important data regularities, we argue that a plot 
of the (log-)size against the rank of the river's streamflow 
(see Figure~\ref{fig:log_rank}) suggests that this relation can be well approximated 
by a discrete generalized Beta distribution (DGBD) of the form
\begin{equation}
f(r) = \frac{A(N + 1 - r)^b}{r^a} 
\end{equation}
where $r$ is the rank, $N := \max(r)$, and $A$, $a$, and $b$ are parameters to be estimated from the data.\footnote{Parameters can be estimated by Ordinary Least Squares from the model in logarithms.} This is indeed the case, and the $R^2$ of the approximating function is $R^2 \approx 0.986$, confirming the ``universality'' of the DGBD suggested in other fields of investigation 
[\cite{martinezmekleretal2009,ponemarcuniv}], particularly in relation to  
complex phenomena characterized by the co-existence of many subsystems
whose interactions produce the observed outcome [\cite{naumiscocho2008}]. 

\begin{table}
\begin{center}
\begin{tabular}{rr}
\hline \hline
$m^3/s$& freq \\ \hline
5.9833 & 10061 \\
5.6859 & 10014 \\
6.2890 &  9940 \\
5.3968 &  9086 \\
6.6030 &  8263 \\
5.1161 &  7978 \\
4.8436 &  6784 \\
6.9253 &  6698 \\
7.9420 &  6361 \\
7.5948 &  5881 \\ \hline \hline
\end{tabular}
\end{center}
\vspace{-0.5cm}\caption{Frequencies of the ten most common values.}
\label{tab:frequencies}
\end{table}

\section{Conclusions}\label{sec:conclusion}

This paper presents the analysis of the Paglia river streamflow.
The dataset used is of large size: the considered series has
30-minutes period over a time-span of more than 22 years. The data series 
is antipersistent, with an average Hurst exponent of about $1/3$.
This gives a precise information on how one can do forecast on the
overflow of the considered river. Furthermore, we show that 
in recent years there has been an increasing trend in
the maxima of the river streamflows, and a periodicity which 
broadly suggests a connection of the streamflows with El Ni\~no 
and the North Atlantic Oscillations. However, more accurate measurements of the streamflow
could be useful in having a more consistent dataset. We also show that
the Paglia river streamflow is well represented by a discrete generalized beta
distribution, congruent with the observed river discharge being the result of
the interaction of many complex subsystems.

\section*{Acknowledgements}
We would like to thank three anonymous referees for their constructive
comments that helped us to substantially improve the paper upon the
previous versions.  
We are pleased to thank Guido Calenda and Elena Volpi for having provided us with the data. 
Funding from the Italian Ministry of Education, University and Research 
(PRIN grant code 201002AXKAJ$_-$005) is gratefully acknowledged. 


\end{document}